\documentclass{jpsj2}

\newcommand{\url}[1]{{\tt#1}}

\begin{document}

\title{Proposal for an interference experiment to test the applicability of quantum theory to
event-based processes\footnote{Accepted for publication in J. Phys. Soc. Jpn.}}

\author{
\textsc{Kristel Michielsen}$^{1}$\thanks{E-mail: k.michielsen@fz-juelich.de},
\textsc{Th. Lippert}$^1$\thanks{E-mail: th.lippert@fz-juelich.de},
\textsc{M. Richter}$^1$\thanks{E-mail: m.richter@fz-juelich.de},
\textsc{B. Barbara}$^2$\thanks{bernard.barbara@grenoble.cnrs.fr},
\textsc{Seiji Miyashita}$^{3}$\thanks{E-mail: miya@spin.phys.s.u-tokyo.ac.jp},
and
\textsc{Hans De Raedt}$^{4}$\thanks{E-mail: h.a.de.raedt@rug.nl},
}%

\inst{
$^1$ Institute for Advanced Simulation, J\"ulich Supercomputing Centre,
  Research Centre J\"ulich, D-52425 J\"ulich, Germany
  \\
$^2$ Institut N\'eel, CNRS, 25 Ave. des Martyrs, BP 166, 38 042 Grenoble Cedex 09, France
\\
$^3$ Department of Physics, Graduate School of Science, The University of Tokyo, 7-3-1 Hongo, Bunkyo-Ku, Tokyo 113-8656, Japan \\
and
  CREST, JST, 4-1-8 Honcho Kawaguchi, Saitama 332-0012, Japan
\\
$^4$ Department of Applied Physics, Zernike Institute for Advanced Materials,
  University of Groningen, Nijenborgh 4, NL-9747 AG Groningen, The Netherlands
}%

\abst{%
We analyze a single-particle Mach-Zehnder interferometer experiment
in which the path length of one arm may change (randomly or systematically) according to the value
of an external two-valued variable $x$, for each passage of a particle through the interferometer.
Quantum theory predicts an interference pattern that is independent of the sequence
of the values of $x$.
On the other hand, corpuscular models that reproduce the results of quantum optics experiments carried out up to
this date show a reduced visibility and a shift of the interference pattern depending on
the details of the sequence of the values of $x$.
The proposed experiment will show that:
(1) it can be described by quantum theory, and thus not by the current corpuscular models,
or
(2) it cannot be described by quantum theory but can be described by the corpuscular models or variations thereof,
or
(3) it can neither be described by quantum theory nor by corpuscular models.
Therefore, the proposed experiment can be used to determine to what extent
quantum theory provides a description of observed events beyond the usual statistical level.
}
\kword{Quantum Mechanics, interference, Mach-Zehnder interferometer}

\maketitle

\section{Introduction}
\subsection{Mach-Zehnder interferometer with light}
\begin{figure}[t]
\begin{center}
\includegraphics[width=14cm]{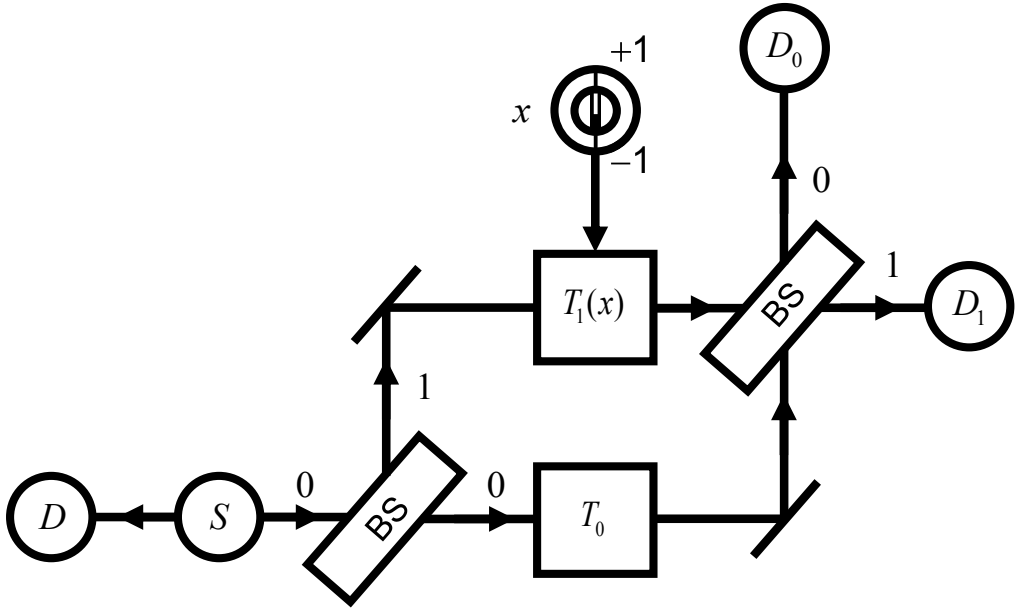}
\caption{Schematic diagram of the proposed Mach-Zehnder interferometer (MZI) experiment.
S: Light source;
BS: 50-50 beam splitter;
$T_0$: Fixed time-of-flight;
$T_1(x)$: Variable time-of-flight
controlled by the external variable $x$;
$D$, $D_{0}$, $D_{1}$: Detectors.
In single photon experiments $x$ may change before the photon enters the MZI but not
during the passage of the photon through the MZI.
For simplicity we consider experiments in which $x$ takes the values $-1$ and $+1$ only.
The recorded dataset for $N$ detection events is given by $\{x_i,d_{0,i},d_{1,i},d_i| i=1,\ldots N\}$
where $d_{k,i}=1$ if detector $D_k$, $k=0,1$ fired and $d_{k,i}=0$ otherwise,
and $d_i=1$ ($d_i=0$) if detector $D$ fired (did not fire).
Note that the value of the experimental setting parameter $x$ is not measured but is known and certain
at each moment in time.
}
\label{fig1}
\end{center}
\end{figure}

Consider the schematic diagram (Fig.~\ref{fig1}) of the Mach-Zehnder interferometer (MZI) experiment in
which the length of the upper arm can be varied by a control parameter $x$, that is $L_1(x)=cT_1(x)$ where $c$
denotes the speed of light and $T_1(x)$ the variable time-of-flight controlled by the variable $x$.
The length of the lower arm is fixed and is given by $L_0=cT_0$.
According to Maxwell's theory, carrying out the experiment with a fixed value of $x$ and with a coherent
monochromatic light source $S$ with frequency $\omega$ gives for the normalized intensities $I_0$ and $I_1$,
recorded by the detectors $D_0$ and $D_1$,~\cite{BORN64}
\begin{eqnarray}
I_{0}&=&\sin^2\frac{\omega(T_0-T_1(x))}{2}=\sin^2\frac{\phi_0-\phi_1(x)}{2}
,
\label{mzi0a}
\\
I_{1}&=&\cos^2\frac{\omega(T_0-T_1(x))}{2}=\cos^2\frac{\phi_0-\phi_1(x)}{2}
,
\label{mzi0b}
\end{eqnarray}
where $\phi_0=\omega T_0$ and $\phi_1(x)=\omega T_1(x)$.
Equations~(\ref{mzi0a}) and (\ref{mzi0b}) show that the signal on the detectors $D_0$ and $D_1$, respectively,
is modulated by the difference between
the time-of-flights $T_0$ and $T_1(x)$ in the lower and upper arm
of the interferometer, respectively, or in other words by the phase difference $\phi_0-\phi_1(x)$.

\subsection{Mach-Zehnder interferometer with single photons: Evidence for the particle nature of photons}
Replacing the light source by a source emitting pairs of single photons yields the same interference
patterns Eqs.~(\ref{mzi0a}) and (\ref{mzi0b}), as shown in a laboratory experiment by Grangier {\sl et al.}~\cite{GRAN86}.
Grangier {\it et al.}~\cite{GRAN86} performed two types of experiments, one with the setup of a MZI and one in which the second
beam splitter has been removed.
A key feature in their experiment is the use of the three-level cascade photon emission of the calcium atom.
When the calcium atoms are excited to the third lowest level, they relax to the second lowest state, emitting photons of frequency
$f$, followed by another transition to the ground state level causing photons of frequency $f'$ to be emitted~\cite{KOCH67}.
It is observed that each such two-step process emits two photons in two spatially well-separated directions,
allowing for the cascade emission to be detected using a time-coincidence technique~\cite{KOCH67}.
One of two light beams produced by the cascade is directed to detector $D$.
The other beam is sent through a 50-50 beam splitter to detectors $D_0$ and $D_1$.
Time-coincidence logic is used to establish the emission of the photons by the three level cascade process:
Only if detectors $D$ and $D_0$, $D$ and $D_1$, or $D_0$ and $D_1$ fire, a cascade emission event occurred.
Then, the absence of a coincidence between the firing of detectors $D_0$ and $D_1$
provides unambiguous evidence that the photon created in the cascade and passing through the beam splitter
behaves as one indivisible entity.
The analysis of the experimental data strongly supports the hypothesis that
the photons created by the cascade process in the calcium atom are to be regarded as indivisible~\cite{GARR09}.

Having established the corpuscular nature of single photons,
Grangier {\sl et al.}~\cite{GRAN86} extend the experiment
by sending the photons emerging from the beam splitter to another beam splitter,
thereby constructing a MZI.
Note that the mirrors do not alter the indivisible character of the photons,
and that the removal of the second beam splitter yields an
experimental configuration that is equivalent to the one used to demonstrate
the corpuscular nature of single photons.
With the second beam splitter in place, Grangier {\sl et al.}~\cite{GRAN86}
observe that after collecting many photons one-by-one, the normalized frequency distributions of detection counts
recorded by the detectors $D_0$ and $D_1$
fit nicely to the interference patterns Eqs.~(\ref{mzi0a}) and (\ref{mzi0b}):
What they observe is the same result as if the source would have emitted a wave.
Thus, it is shown that one-by-one, these photons build up an interference pattern.

\subsection{Theoretical description of single-photon Mach-Zehnder interferometer}
If we want to use classical concepts (Newtonian mechanics and classical electrodynamics) to interpret the results of these single-photon experiments,
one must use a particle picture for the experiment in which the second beam splitter has been removed
and a wave picture to explain the observation of interference when the second beam splitter is left in place.
A question that is then often posed is whether photons are particles or waves, because classically they cannot be both.

Quantum theory resolves this dissension by introducing the notion of particle-wave duality
and complementarity, in the sense that different experiments are required to observe the particle or wave property of photons but that they possess both.
However, the three above mentioned theories are silent about the underlying physical processes
by which single particles, arriving one-by-one at a detector gradually build up an interference pattern.

\subsection{Event-based corpuscular model}
In Refs.~\cite{RAED05d,RAED05b,MICH11a} we have proposed an event-based corpuscular model, see section 4 for a short description,  which has shown to reproduce the statistical predictions of quantum theory for the
single beam splitter and the MZI experiment of Grangier {\sl et al.}~\cite{GRAN86}.
For the latter experiment, the event-based corpuscular model allows for a particle-only description of
the interference pattern.
In a pictorial description of the simulation model, we may speak about ``photons'' generating the detection events. However,
these so-called photons are elements of a model or theory for the real laboratory
experiment only. The experimental facts are the settings of the various optical apparatuses and
the detection events. What happens in between activating the source and the registration
of the detection events is not measured and is therefore not known. Although in the event-based model one always
has full which-path-information of the individual photons (one can always track the photons during the
simulation), the photons build up an interference pattern at the detector. Hence, although, the
appearance of an interference pattern is commonly considered to be characteristic for
a wave, we have demonstrated that, as in experiment, it can also be build up by many photons~\cite{RAED05d,RAED05b,MICH11a}.
Thus, in contrast to the quantum theoretical description of the MZI experiment in terms of averages over many events, the event-based corpuscular model provides
a rational, logically consistent explanation of the experimental facts in terms of causal processes that are formulated as
discrete events to which one can associate ``particles''.

Using the same algorithmic approach for modeling the single beam splitter and MZI experiment with
single photons of Grangier {\sl et al.}~\cite{GRAN86} (see Refs.~\cite{RAED05d,RAED05b,MICH11a}),
we also modeled Wheeler's delayed choice experiment with single photons of Jacques {\sl et al.}~\cite{JACQ07}
(see Refs.~\cite{ZHAO08b,MICH10a,MICH11a}),
the quantum eraser experiment of Schwindt {\sl et al.}~\cite{SCHW99} (see Ref.~\cite{JIN10c,MICH11a}),
double-slit and two-beam single-photon interference experiments and the single-photon interference experiment with
a Fresnel biprism of Jacques {\sl et al.}~\cite{JACQ05} (see Ref.~\cite{JIN10b,MICH11a}),
quantum cryptography protocols (see Ref.~\cite{ZHAO08a}),
the Hanbury Brown-Twiss experiment of Agafonov {\sl et al.}~\cite{AGAF08} (see Ref.~\cite{JIN10a,MICH11a}),
universal quantum computation (see Ref.~\cite{RAED05c,MICH05}),
the violation of Bell's inequalities in Einstein-Podolsky-Rosen-Bohm-type of experiments, involving two photons in the singlet state, of Aspect {\sl et al.}~\cite{ASPE82a,ASPE82b}
and Weihs {\sl et al.}~\cite{WEIH98} (see Refs.~\cite{RAED06c,RAED07a,RAED07b,RAED07c,RAED07d,ZHAO08,MICH11a}),
and the propagation of electromagnetic plane waves through homogeneous thin films and stratified media (see Ref.~\cite{TRIE11,MICH11a}).
A review of the simulation method and its applications is given in Ref.~\cite{MICH11a}.
Interactive demonstration programs, including source codes, for some of the single-photon experiments
are available for download~\cite{COMPPHYS,MZI08,DS08}.
A computer program to simulate single-photon Einstein-Podolsky-Rosen-Bohm experiments that violate Bell's inequality can be found in Ref.~\cite{RAED07b}.

For many different optics experiments
the event-based corpuscular model reproduces the probability distributions of quantum theory or results of Maxwell's wave theory by assuming that photons have a particle-character
only.
The event-based corpuscular model is free of paradoxes that result from the assumption that photons exhibit dual, wave/particle behavior.
A crucial property of the event-based corpuscular models is that they reproduce these ``wave results'' observed in different experiments without any change
to algorithms modeling the photons and optical apparatuses~\cite{MICH11a}.
These algorithms can, of course, be simplified
for particular experiments. For example, if photon polarization is not essential to a given experiment, then for
simplicity we can omit the photon polarization in the event-based corpuscular model of this particular experiment.

Although these algorithms can be given an interpretation as a realistic cause-and-effect description that
is free of logical difficulties, it is at present impossible to decide whether or not such algorithms are realized by Nature:
Only new, dedicated experiments such as the one proposed in this paper can teach us more about this intriguing question.

\subsection{Applicability of quantum theoretical and corpuscular model descriptions of single-particle experiments}
Given the fact that the frequency distributions produced by the event-based corpuscular models cannot be distinguished from those predicted by quantum theory
for the single-photon experiments performed so far and given the general belief that quantum theory can be used to describe all single-particle experiments, the key question
is whether an experiment can be performed that shows a difference between the results obtained by quantum theory and those obtained by the event-based corpuscular model for this experiment.
A trivial idea, that however cannot be realized in the laboratory, is to compare one MZI experiment with $N$ photons passing through it with
$N$ identical MZI experiments in which exactly one photon passes through each of the $N$ MZIs.
Quantum theory predicts that after collecting the $N$ single-photon detection events in both cases the same interference pattern is obtained.
However, the question whether or not for this particular example quantum theory describes what will be observed in the laboratory will remain unanswered:
In the laboratory, it is very difficult, not to say practically impossible, to perform $N$ identical experiments  with $N$ sufficiently large.
On the other hand, it is clear that the event-based corpuscular model~\cite{RAED05d,RAED05b,MICH11a} will not give identical results for both cases: Only in the first case will it reproduce the
same interference pattern as observed in the laboratory experiment and as given by quantum theory.
However, one obviously cannot refute a model on the basis of imagined outcomes of an experiment that
is practically impossible to realize in the laboratory.

In Refs.~\cite{RAED05d,RAED05b,MICH11a} we have shown that the event-based corpuscular model can produce frequency distributions that cannot be distinguished from
those predicted by quantum theory for single photon MZI experiments that are performed in the stationary regime, that is under experimental
conditions that remain the same for a relatively large number of incoming photons.
For this experiment quantum theory predicts the same interference pattern independent of the number of incoming photons while for
the event-based corpuscular model one can recognize a transient and a stationary regime, only the latter one giving rise to the same
interference pattern as the one predicted by quantum theory. In the quantum theoretical description there is no transient regime.

The analysis of the transient regime in a laboratory single-photon experiment might be extremely difficult because most experiments
require a stabilization procedure. Therefore, in most cases the first experimental recordings are discarded in the analysis.
In order to study the transient effects and their eventual importance we propose the modified Mach-Zehnder experiment
described in this paper. The question to answer is whether the transient effects can be observed experimentally or not.
If the experiment shows transient effects in the interference pattern this would indicate that quantum theory cannot be
used to describe the outcome of this modified MZI experiment.

\subsection{Proposal for a single-particle interference experiment to test the applicability of quantum theory and corpuscular models}\label{prop}
In this paper we present an analysis of a single-photon interference experiment that can be
modeled in terms of particles only, that can be experimentally tested and
for which the event-based corpuscular models predict that for some experimental conditions,
the results differ from those predicted by quantum theory.
Hence, under these conditions there are three possible conclusions:
\begin{itemize}
\item{the experiment can be described by quantum theory, and thus not by the current event-based corpuscular models,}
\item{the experiment cannot be described by quantum theory but can be described by the current event-based corpuscular models or variations thereof,}
\item{the experiment can neither be described by
quantum theory nor by the current event-based corpuscular models.}
\end{itemize}

Specifically, we consider the MZI experiment (see Fig.~\ref{fig1}) in which we allow the variable $x$ to
change before the particle enters the MZI but not
during the passage of the particle through the MZI.
Note that the value of $x$ is not measured but is always known and certain.
In sections 5 and 7 we demonstrate that this experiment may be
used to determine the conditions under which quantum theory fails to describe single particle detection events
or to refute the event-based corpuscular model proposed in Refs.~\cite{RAED05d,MICH11a} which has shown
to reproduce the statistical predictions of quantum theory
for the MZI experiment with fixed $x$.

For simplicity, but not out of necessity, we only consider experiments in which $x$ takes the values
+1 and -1 and for which
$\phi_1(x=+1) \mathrm{mod} 2\pi=0$ and $\phi_1(x=-1)  \mathrm{mod} 2\pi=-\pi/2$.
We consider a systematic and a random procedure to change $x$ such that $x=+1$ and $x=-1$ occur
with the same frequency.
In the systematic procedure we replace $x$ by $-x$ after the single photon source has emitted $K$ photons.
For $K=1$ this procedure leads to an alternating sequence of $x$-values.
In the random procedure we use a random number to decide whether or not we replace $x$ by $-x$ after the single
photon source has emitted $K$ photons.
In both procedures we repeat this sequence so that the total number of photons emitted by the source equals $N$.
Each click of the detector $D_0$ or $D_1$ is labeled by the currently known and certain value of $x$.
We do not allow for an ``on purpose'' mislabeling of the detection events by a wrong value of $x$.
After the $N$ photons have been sent and all clicks have been registered,
we count the number of detection events on $D_0$ and $D_1$ for each value of $x$
separately, yielding the numbers $N_0(x)$ and $N_1(x)$.
Finally, we define the normalized frequencies to
detect photons by $F_0(x)=N_0(x)/(N_0(x)+N_1(x))$ and $F_1(x)=N_1(x)/(N_0(x)+N_1(x))$.
Note that we made no assumption about the detection efficiency (see section 7).

\section{Impact of the proposed experiment}
In contrast to the event-based corpuscular model (see section 4),
quantum theory predicts results that are independent of the sequence of $x$-values (see section 3).
In quantum theory, the probability wave propagates simultaneously along both arms of the MZI.
Therefore, for each value of $x$ quantum theory predicts the corresponding interference pattern, independent of the sequence of $x$-values.
At this moment there is no experimental test of this independence.
Therefore it is worthwhile to perform an experiment as described in section 7, even
if the general belief would be that quantum theory correctly describes the experiment
and predicts results independent of the sequence of $x$-values.

\section{Quantum theory}

According to wave theory~\cite{BORN64}, the amplitudes ($b_0(x),b_1(x))$
of the photons in the output modes 0 and 1 of the
MZI with a fixed value of $x$ are given by
\begin{equation}
\left(
\begin{array}{c}
b_0(x)\\
b_1(x)
\end{array}
\right)
=ie^{i\varphi^{\prime}(x)}
\left(
\begin{array}{cc}
\sin\varphi(x)&\cos\varphi(x)\\
\cos\varphi(x)&-\sin\varphi(x)
\end{array}
\right)
\left(
\begin{array}{c}
a_0\\
a_1
\end{array}
\right)
,
\label{mzi1}
\end{equation}
where the amplitudes of the photons in the input modes 0 or 1 are represented
by $a_0$ and $a_1$, $\varphi(x)=(\phi_0-\phi_1(x))/2$
and $\varphi^{\prime}(x)=(\phi_0+\phi_1(x))/2$.
For the case at hand $a_1=0$ and without loss of generality, we may take
$a_0=1$.

The Copenhagen interpretation maintains that the wave function provides a complete
and exhaustive description of the experiment with an individual particle~\cite{HOME97,BALL03}.
Therefore, grouping all detection events of the individual photons according
to the corresponding values of $x$ at the time of their passage through the MZI,
the Copenhagen interpretation predicts that the probability distributions to register detection events at $D_0$ are given by
\begin{eqnarray}
I_{0}(x=+1)&=&|b_0(x=+1)|^2=\frac{1}{2}\sin^2\frac{\phi_0}{2}
,
\label{mzi2a}
\\
I_{0}(x=-1)&=&|b_0(x=-1)|^2=\frac{1}{2}\sin^2\frac{\phi_0+\pi/2}{2}
,
\label{mzi2b}
\end{eqnarray}
where the prefactor 1/2 comes from the fact that we have assumed that $x=+1$ and $x=-1$ occur with the
same frequency. Note that
Eqs.~(\ref{mzi2a}) and (\ref{mzi2b}) are independent of the procedure that changes $x$.

If the detection events are not grouped according to the values of $x$, the Copenhagen interpretation predicts
\begin{equation}
I_0^{\prime}=\frac{1}{2}\sin^2\frac{\phi_0}{2}+\frac{1}{2}\sin^2\frac{\phi_0+\pi/2}{2}.
\label{mzi2c}
\end{equation}
Here and in the following the prime indicates that the detection events are not grouped (associated) with the
current value of $x$ at the time of detection.

Finally, if $x$ does not change during the experiment
\begin{eqnarray}
I_0^{\prime\prime}(x=+1)&=&\sin^2\frac{\phi_0}{2},
\label{mzi2d}
\\
I_0^{\prime\prime}(x=-1)&=&\sin^2\frac{\phi_0+\pi/2}{2},
\label{mzi2e}
\end{eqnarray}
where the double prime indicates that the value of $x$ is fixed during the experiment.

The statistical interpretation maintains to provide a description of the statistical properties
of an ensemble of similarly prepared systems only~\cite{BALL03}.
For the case at hand, the output state of the MZI is represented by the density matrix
$\widehat\rho=\sum_{y=\pm 1}\rho (y)$, where
\begin{eqnarray}
\rho (y)=
\left(
\begin{array}{cc}
b_0^*(y)b_0(y)&b_0^*(y)b_1(y)\\
b_1^*(y)b_0(y)&b_1^*(y)b_1(y)
\end{array}
\right)
.
\label{mzi3}
\end{eqnarray}
The probability to register detection events in output channel 0 of the MZI,
is given by $I_0(x)=\sum_{y=\pm 1}\mathop{\bf Tr}\rho (y){\widehat I}_0(x,y)$ where
\begin{eqnarray}
{\widehat I}_0(x,y) =
\left(\begin{array}{cc}
1& 0\\
0& 0
\end{array}
\right)\delta_{x,y}
,
\label{mzi4}
\end{eqnarray}
also yielding Eqs.~(\ref{mzi2a})-(\ref{mzi2e}).

Both the Copenhagen and the statistical interpretation predict the same outcome for the proposed experiment,
as it should be because the mathematical formalism of quantum theory itself is free of interpretation.
Note that the quantum theoretical description of the experiment would be different if
$x$ is not considered to be a parameter of the experimental setting which is known and certain
at every moment in time, but is part of the measurement outcome and considered to be unknown until measured.

Taken literally, one may think that even for one particle the Copenhagen interpretation predicts an interference pattern
but this contradicts the experiment in which only one click, either of $D_0$ or of $D_1$, is registered.
This apparent contradiction is a manifestation of the quantum measurement paradox:
Although quantum theory provides a recipe to compute the frequencies for observing events,
it does not describe individual events, such as
the arrival of a single electron at a particular position on the screen or the detection
of a single photon by a particular detector~\cite{HOME97}.
The statistical interpretation tactically avoids the measurement paradox by being silent on the issue of individual events.

If quantum theory correctly describes the experiment with varying but always known $x$,
we expect to find for the observed frequencies at detector $D_0$
\begin{eqnarray}
F_{0}(x=+1)&\approx&I_{0}(x=+1)=\frac{1}{2}\sin^2\frac{\phi_0}{2}
,
\label{mzi5a}
\\
F_{0}(x=-1)&\approx&I_{0}(x=-1)=\frac{1}{2}\sin^2\frac{\phi_0+\pi/2}{2}
,
\label{mzi5b}
\end{eqnarray}
(see Eqs.~(\ref{mzi2a}) and (\ref{mzi2b})) independent of the procedure for changing $x$ being
systematic or random and
independent of the number of emitted photons $K$ per change of $x$.
In fact, quantum theory predicts that the result is completely independent of the sequence of $x$.
Note that this cannot be true in general:
One could consider $x=+1,\dots,+1,-1$ so that there is only one
event for $x=-1$.
In this case the observed frequency does not correspond to an interference pattern although quantum theory predicts that also for this case
$I_0(x=-1)=\sin^2 (\phi_0+\pi/2)/2$.

It is precisely this feature, the fact that quantum theory predicts results that are independent of
the sequence of $x$-values, that we propose to test experimentally.
Note that there is no indication, let alone a kind of proof that quantum theory, being a theory that makes predictions about statistics only,
correctly describes experiments in which the procedure for preparing
the state of the photon (i.e. the state before the photon is being detected)
can change with each photon.

\section{Corpuscular model for interference}

Although detailed accounts of the event-based corpuscular modeling approach, with applications to many different
single-photon experiments have been published elsewhere~\cite{RAED05d,RAED05b,RAED05c,MICH05,RAED06c,%
RAED07a,RAED07b,RAED07c,RAED07d,ZHAO08a,ZHAO08,ZHAO08b,MICH10a,JIN10a,JIN10b,JIN10c,MICH11a},
for the reader's convenience, we briefly describe the simulation technique.
The basic ideas of the simulation approach are that (i) we stick to what we know about the experiment, that is we consider the experimental configuration and its outcome
as input for constructing the simulation algorithm; (ii) we try to invent a procedure, algorithm or set of rules that generates the same type of data as in experiment and
reproduces the averages predicted by quantum theory; (iii) we keep compatibility with macroscopic concepts.

Generally speaking, the event-based corpuscular simulation method can be viewed as a message passing and message processing method in which the photons play the role of the messengers
and the optical apparatuses, such as a (polarizing) beam splitter, polarizer, wave plate, detector and so on play the role of the processors that interpret and manipulate the messages.
In what follows we briefly describe how we model the photon and the optical apparatuses that are sufficient to simulate a MZI experiment.
This means that here we do not consider the polarization of the photon and that we consider detectors that simply count the detection events.
More sophisticated models for the photon and the detectors can be found in
Refs.~\cite{MICH11a,RAED06c,RAED07a,RAED07b,RAED07c,ZHAO08,ZHAO08a,ZHAO08b,JIN10c,MICH10a}
and \cite{JIN10a,JIN10b,MICH11a}, respectively.
Note that these more sophisticated event-based corpuscular models have also been used to simulate the MZI experiment.
They would, however, unnecessarily complicate the modeling and pictorial description of the experiments we consider here.
To simulate the MZI experiment we make use of the following models:
\begin{itemize}
\item{Photon:}{
We consider the photon to be a particle having an internal clock with one hand that rotates with a frequency $f=\omega /2\pi$.
Hence, the rotation velocity of the hand depends on the angular frequency $\omega$, that is the ``color'' of the photon.
Thus, the hand of a blue photon rotates faster than the hand of a red photon.
As the photon travels from one position in space to another, the clock encodes its time of flight $t$ modulo the period $1/f$.
We therefore view the photon as a messenger carrying as message the position of the clock's hand.
We encode the message as a two-dimensional unit vector ${\bf e}=(e_0,e_1)=(\cos\omega t, \sin\omega t)$.
This particle model for the photon was previously used by Feynman in his theory of quantum electrodynamics~\cite{FEYN85}.
Feynman used the position of the clock's hand to calculate the probability amplitudes.
Although quantum electrodynamics resolves the wave-particle duality by saying that light is made of particles (as Newton originally thought), it is
only able to calculate the probability that a photon will hit a detector, without offering a mechanism of how this actually happens~\cite{FEYN85}.
}
\item{Source:}{
The source creates a messenger (photon), carrying a message as described above, and waits until its message has been processed by a detector before creating the next messenger.
Hence, there can be no direct communication between the messengers. Therefore, the simulation model (trivially) satisfies Einstein's criterion of local causality.
When a messenger is created, its internal clock time is set to zero.
We label the messengers and their messages by a subscript $n\ge 0$.
}
\item{Beam splitter:}{
The processor modeling a beam splitter consists of three stages: An input stage, a transformation stage and an output stage.
The input stage has two input channels labeled by $k=0,1$, two registers ${\bf Y}_k = (Y_{0,k}, Y_{1,k})$ and an
internal two-dimensional vector ${\bf u} = (u_0, u_1)$ with the additional
constraints that $u_i\ge0$ for $i=0, 1$ and that $u_0 + u_1 = 1$.
The $(n+1)$-th messenger carrying the message ${\bf e}_{n+1}=(e_{0,n+1},e_{1,n+1})$ arrives at input channel 0 or input channel 1.
If the messenger arrives on input channel 0 (1), then register ${\bf Y}_0$ (${\bf Y}_1$) stores the message brought by the messenger, that is ${\bf Y}_0 = (e_{0,n+1},e_{1,n+1})$ (${\bf Y}_1 = (e_{0,n+1},e_{1,n+1})$).
Note that only one of the two registers is updated when a messenger arrives at the processor.
After arrival of  the $(n+1)$-th  messenger on input channel $k=0,1$ the input stage also updates its internal vector
according to the rule $u_{i,n+1} = \alpha u_{i,n}+(1-\alpha )\delta_{i,k}$ where $0<\alpha <1$ is a parameter.
The $u_{k,n}$ can be interpreted as (an estimate of) the probability for the arrival of a messenger on input channel $k$ and $\alpha$ can be interpreted as a parameter
controling the learning process of this (estimate of the) probability~\cite{RAED05d,MICH11a}.
}

The transformation stage takes the six values stored in the two registers ${\bf Y}_0$, ${\bf Y}_1$ and the internal vector ${\bf u}$ and transforms this data according to the rule
\begin{eqnarray}
\frac{1}{\sqrt{2}}
\left(
\begin{array}{c}
Y_{0,0}\sqrt{u_0}-Y_{1,1}\sqrt{u_1}\\
Y_{0,1}\sqrt{u_1}+Y_{1,0}\sqrt{u_0}\\
Y_{0,1}\sqrt{u_1}-Y_{1,0}\sqrt{u_0}\\
Y_{0,0}\sqrt{u_0}+Y_{1,1}\sqrt{u_1}
\end{array}
\right)
&{\longleftarrow}&
\left(
\begin{array}{c}
Y_{0,0}\sqrt{u_0}\\
Y_{1,0}\sqrt{u_0}\\
Y_{0,1}\sqrt{u_1}\\
Y_{1,1}\sqrt{u_1}
\end{array}
\right),
\label{BS1}
\end{eqnarray}
where we have omitted the messenger label $(n+1)$ to simplify the notation.
Using two complex numbers instead of four real numbers
Eq.~(\ref{BS1}) can also be written as
\begin{eqnarray}
\frac{1}{\sqrt{2}}
\left(
\begin{array}{c}
Y_{0,0}\sqrt{u_0}-Y_{1,1}\sqrt{u_1}+i(Y_{0,1}\sqrt{u_1}+Y_{1,0}\sqrt{u_0})\\
Y_{0,1}\sqrt{u_1}-Y_{1,0}\sqrt{u_0}+i(Y_{0,0}\sqrt{u_0}+Y_{1,1}\sqrt{u_1})
\end{array}
\right)
\nonumber \\
\hbox to 1cm{}{\longleftarrow}
\left(
\begin{array}{c}
Y_{0,0}\sqrt{u_0}+i Y_{1,0}\sqrt{u_0}\\
Y_{0,1}\sqrt{u_1}+i Y_{1,1}\sqrt{u_1}
\end{array}
\right).
\label{BS2}
\end{eqnarray}
Identifying $a_0$ with $Y_{0,0}\sqrt{u_0}+i Y_{1,0}\sqrt{u_0}$
and $a_1$ with $Y_{0,1}\sqrt{u_1}+i Y_{1,1}\sqrt{u_1}$
it is clear that the transformation Eq.~(\ref{BS2}) plays the
role of the matrix-vector multiplication
\begin{eqnarray}
\left(
\begin{array}{c}
b_0\\
b_1
\end{array}
\right)
=
\frac{1}{\sqrt{2}}
\left(
\begin{array}{c}
a_0+ia_1\\
a_1+ia_0
\end{array}
\right)
=
\frac{1}{\sqrt{2}}
\left(
\begin{array}{cc}
1&i\\
i&1
\end{array}
\right)
\left(
\begin{array}{c}
a_0\\
a_1
\end{array}
\right),
\label{BS3}
\end{eqnarray}
where the $(a_0,a_1)$ ($(b_0,b_1)$) denote the amplitudes
of the photons in the input (output) output modes 0 and 1 of a
beam splitter~\cite{RAED05d,MICH11a}.

The output stage of the processor uses the content of the four-dimensional vector in Eq.~(\ref{BS1}) to update the message
carried by the $(n+1)$th messenger and directs this messenger to one of its two output channels
labeled by $k=0,1$.
The output stage sends the $(n+1)$-th messenger with message
${\bf w}_{n+1}=( Y_{0,0}\sqrt{u_0}-Y_{1,1}\sqrt{u_1},Y_{0,1}\sqrt{u_1}+Y_{1,0}\sqrt{u_0})/\sqrt{2}$
through output channel 0 if
$w_{0,n+1}^2+w_{1,n+1}^2>r$
where $0<r<1$ is a uniform random number.
Otherwise, it sends the message
${\bf z}_{n+1}=( Y_{0,1}\sqrt{u_1}-Y_{1,0}\sqrt{u_0},Y_{0,0}\sqrt{u_0}+Y_{1,1}\sqrt{u_1})/\sqrt{2}$
through output channel 1.

\item{Detector:}{
In the MZI experiment the detectors are counters that simply count the number of messengers (photons) that they receive.
In the modified MZI experiment we propose here, the detectors $D_0$ and $D_1$ each have two counters, one for
counting detection events corresponding to the parameter setting $x=-1$ and one for counting detection events corresponding
to the parameter setting $x=+1$. Hence, in total we have four counters: $N_0(x=-1)$, $N_0(x=+1)$, $N_1(x=-1)$ and $N_1(x=+1)$.
Recall that $x$ is a parameter of which the value is always known with certainty.}
\end{itemize}

\section{Simulation results}
\subsection{Detection events not grouped according to the values of $x$}
\begin{figure}[t]
\begin{center}
\includegraphics[width=14cm]{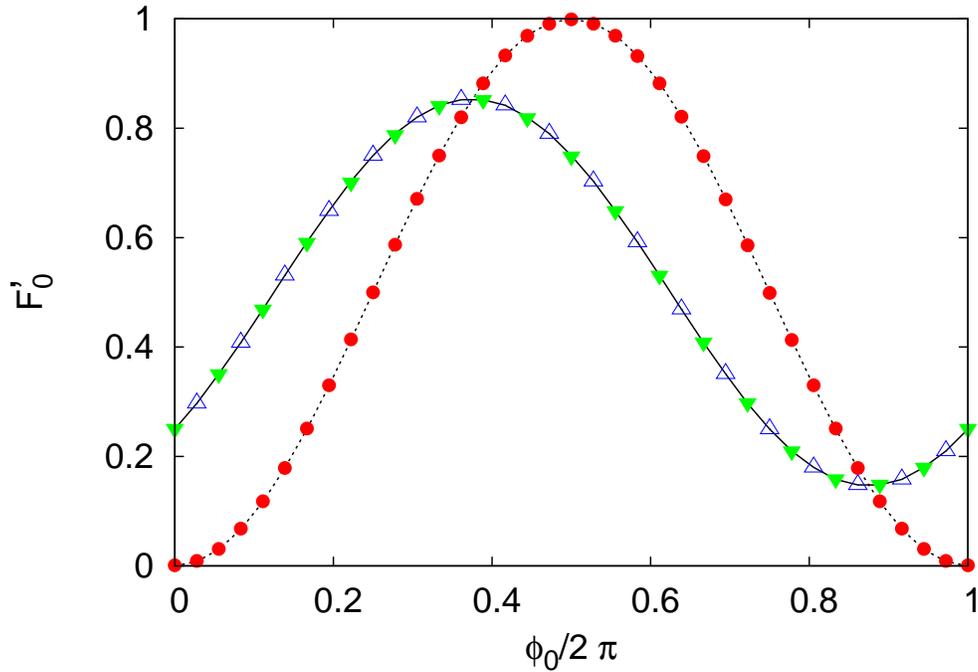}
\caption{%
Results for the normalized frequency $F'_0$ of detection events that are not grouped according to the value of $x$.
Data are obtained from simulations employing
fully classical, locally causal, corpuscular models~\cite{RAED05d,MICH11a} for all the components of the
MZI experiment shown in Fig.~\ref{fig1}.
For each value of $\phi_0$, $N=10^6$ input events
were generated and the model parameter $\alpha=0.99$.
Dotted line: Prediction of quantum theory, see Eq.~(\ref{mzi2d});
Solid line: Prediction of quantum theory, see Eq.~(\ref{mzi2c});
Bullets: Simulation data for $x=+1$ fixed;
Solid triangles:
Simulation data for the case that $x$ changes sign ($x=+1,-1,+1,\ldots$) with each photon emitted,
corresponding to the systematic procedure for changing $x$ with $K=1$;
Open triangles:
Simulation data for the case that $x$ changes sign with every ten photons emitted,
corresponding to the systematic procedure for changing $x$ with $K=10$.
}
\label{fig2}
\end{center}
\end{figure}

In Fig.~\ref{fig2},
we present results for the normalized frequency $F'_0$
as a function of $\phi_0\in [0,2\pi]$ for the experiment in which $x$ is changed according to the
systematic procedure with $K=1,10,N$, where the number of particles $N=10^6$.
The detection events are not grouped (associated) with the value of $x$ at the time of the detection event.
From these data, we see that
\begin{enumerate}
\item{For $K=1$ (solid triangles), that is when $x$ alternates for each photon entering the MZI,
the event-based corpuscular model reproduces the statistical results of quantum theory
(solid line connecting the triangles, as given by Eq.~(\ref{mzi2c})).
}
\item{For $K=10$ (open triangles), that is when $x$ alternates for each ten photons entering the MZI,
there is excellent agreement between the simulation data and the results of quantum theory
(solid line connecting the triangles, as given by Eq.~(\ref{mzi2c})).
}
\item{For $K=N$ (bullets), that is for fixed $x=+1$, $F'_0=F''_0(x=+1)$ and
the event-based corpuscular model reproduces the statistical
results of quantum theory
(dotted line connecting the bullets, as given by Eq.~(\ref{mzi2d})).
}
\end{enumerate}
Summarizing:
For fixed $x$, the results of the event-based corpuscular model are in excellent agreement with Eqs.~(\ref{mzi2d}) and (\ref{mzi2e}), that is with quantum theory.
For varying $x$ and without grouping the detection events according to the value of $x$, the frequencies at detector $D_0$ obtained from
the event-based corpuscular model agree perfectly with the probability distribution Eq.~(\ref{mzi2c}) predicted by quantum theory.
The results do not depend on the number of photons ($K$) per change of $x$.

\subsection{Detection events grouped according to the values of $x$}
Unlike quantum theory, which predicts the probability distributions to be independent of details of the sequence of $x$-values
if the detection events are grouped according to the value of $x$ (see Eqs.~(\ref{mzi2a}) and (\ref{mzi2b})),
the event-based corpuscular model of a MZI makes specific predictions
for the frequencies observed at detector $D_0$
that depend on
the procedure to change $x$ and on
the number of particles $K$ that pass through the MZI
while $x$ is constant.

By construction~\cite{RAED05d,MICH11a}, for fixed $x$ the event-based corpuscular model can produce detection events
with a frequency in perfect agreement with $I_0^{\prime} (x)$ given by Eqs.~(\ref{mzi2d}) and (\ref{mzi2e}) if on average half of the particles
travel along the upper arm of the MZI and half of them along the lower arm. Only in this way the second beam splitter
of the MZI is able to obtain correct information from the particles about the phase difference $\phi_0-\phi_1 (x)$.
If $x$ is changed and if the particle travels along the upper arm, this beam splitter still obtains the correct
information about the changed phase difference. However, if $x$ is changed and if the particle travels along the lower arm,
the particle is unable to pick up the information about the change in path length of the upper arm.
As a result, the second beam splitter of the MZI obtains information about the phase difference which does not
correspond to the value $\phi_0 - \phi_1 (x)$ associated with the new value of $x$.
In this case we say that the resulting detection event is associated with the ``wrong'' value of $x$:
The detection event is generated on the basis of information about a phase difference which does not correspond
to the current value of $x$, which by itself is always known and certain.
From this description it is clear that the effect of these detection events on the observed frequencies after many
detection events have been recorded, strongly depend on the value of $K$: If the number of changes in $x$ is
small compared to the number of photons $N$ ($K\rightarrow N$) then the effect becomes negligible.

From the description of the event-based corpuscular model, it follows directly that
the observed frequencies at detector $D_0$ are given by
\begin{equation}
{\widetilde I}_{0}(x=+1)=\frac{1-E}{2}\sin^2\frac{\phi_0}{2}+\frac{E}{2}\sin^2\frac{\phi_0+\pi/2}{2}
,
\label{mzi6a}
\end{equation}
\begin{equation}
{\widetilde I}_{0}(x=-1)=\frac{1-E}{2}\sin^2\frac{\phi_0+\pi/2}{2}+\frac{E}{2}\sin^2\frac{\phi_0}{2}
,
\label{mzi6b}
\end{equation}
where $0\le E\le1$ is the rate of making wrong associations.
Numerical experiments show that $E\approx 1/(2+2K)$ provides a simple, fairly accurate approximation
of the rate if the random procedure to change $x$ is used. If we would not group the detection events
according to the values of $x$ the observed frequencies at detector $D_0$ would simply be given by
the sum of Eqs.~(\ref{mzi6a}) and (\ref{mzi6b}), what is exactly what we described in the previous section.

Note that the notion of "wrong association" only makes sense in the event-based model because in
this model we can track individual particles when they traverse through the MZI.
In contrast, in the proposed experiment (see section 7), it is impossible for the experimenter
to make "wrong associations" because as described in section 1.6 we do not allow the experimenter
to mislabel on purpose the detection events by a wrong value of $x$.  Similarly, there are no
"wrong associations" in wave theory because the wave propagates simultaneously along both arms
of the interferometer and therefore always carries the correct information about the current setting of $x$.

\begin{figure}[t]
\begin{center}
\includegraphics[width=14cm]{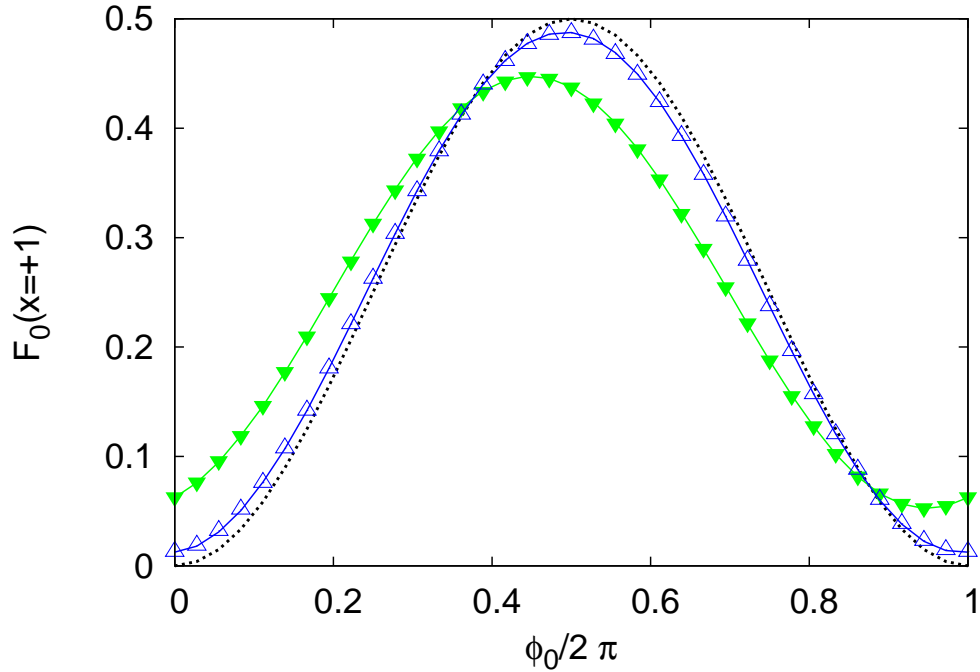}
\caption{%
Results for the normalized frequency $F_0(x=+1)$ of detection events that are grouped according to the value of $x$.
Data are obtained from simulations employing
fully classical, locally causal, corpuscular models~\cite{RAED05d,MICH11a} for all the components of the
MZI experiment shown in Fig.~\ref{fig1}.
For each value of $\phi_0$, $N=10^6$ input events
were generated and the model parameter $\alpha=0.99$.
Dotted line: Prediction of quantum theory, see Eq.~(\ref{mzi2a});
Solid triangles:
Simulation data for the case that $x$ changes sign ($x=+1,-1,+1,\ldots$) with each photon emitted,
corresponding to the systematic procedure for changing $x$ with $K=1$.
The solid line through the data points is given by Eq.~(\ref{mzi6a}) with $E=0.333$.
Open triangles:
Simulation data for the case that $x$ changes sign with every ten photons emitted,
corresponding to the systematic procedure for changing $x$ with $K=10$.
The solid line through the data points is given by Eq.~(\ref{mzi6a}) with $E=0.100$.
}
\label{fig3}
\end{center}
\end{figure}

\begin{figure}[t]
\begin{center}
\includegraphics[width=14cm]{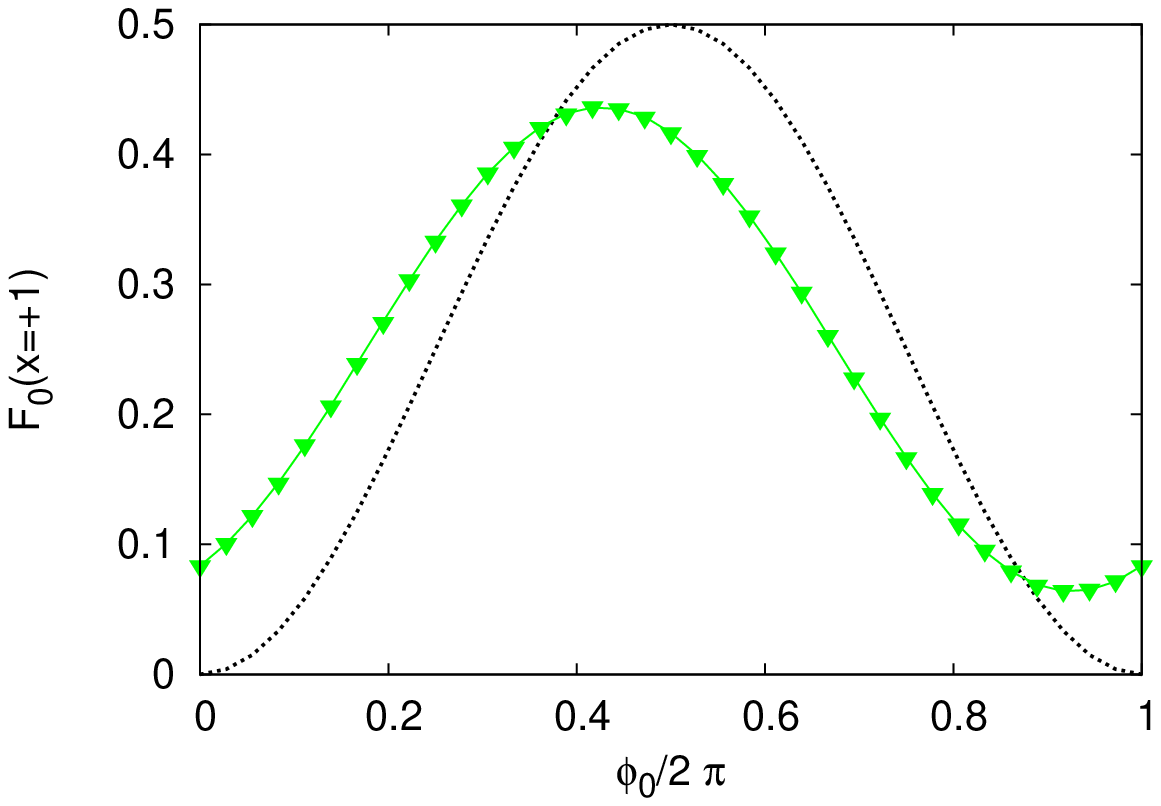}
\caption{%
Same as Fig.~\ref{fig3} except that
$x$ is changed according to the random procedure with $K=1$
and that the solid line through the triangles is given by
Eq.~(\ref{mzi6a}) with $E=1/(2+2K)=1/4$.
}
\label{fig4}
\end{center}
\end{figure}

In Fig.~\ref{fig3},
we present results for the normalized frequency $F_0(x=+1)$
as a function of $\phi_0\in [0,2\pi]$ for the experiment in which $x$ is changed according to the
systematic procedure with $K=1,10$, where the number of particles $N=10^6$.
The detection events are grouped (associated) with the value of $x$ at the time of the detection event.
From these data, we conclude that
\begin{enumerate}
\item{For $K=1$ (solid triangles), that is when $x$ alternates for each photon entering the MZI,
the event-based corpuscular model predicts significant deviations from the results of quantum theory (dotted line, Eq.~(\ref{mzi2a})).
There is excellent agreement between the simulation data and Eq.~(\ref{mzi6a}) (solid line through the solid triangles)
with $E=0.333$.
}
\item{For $K=10$ (open triangles), that is when $x$ alternates for each ten photons entering the MZI,
the difference between the data generated by the event-based corpuscular model and the results of quantum theory (dotted line, Eq.~(\ref{mzi2a}))
becomes rather small.
There is excellent agreement between the simulation data and Eq.~(\ref{mzi6a}) with $E=0.100$
(solid line through the open triangles).
}
\end{enumerate}
Simulations (data not shown) confirm the intuitively evident expectation that as
the number of photons $K$ between changes of $x$ increases, the data produced
by the event-based corpuscular model converge to the prediction of quantum theory Eq.~(\ref{mzi2a}).
This also follows directly from the analytic expression Eq.~(\ref{mzi6a})
because $E\rightarrow0$ if $K\rightarrow N$.

In Fig.~\ref{fig4}, we present simulation data for the case in which
$x$ is changed according to the random procedure with $K=1$.
Qualitatively, the results are the same as when $x$ changes systematically (see Fig.~\ref{fig3}).
However, the rate $E$ is different. For $K=1$, $E=0.333$ for the systematic procedure and $E=0.25$
for the random procedure.
In the case of the random procedure, simulation data for various $K$ (not shown)
are rather accurately represented by Eq.~(\ref{mzi6a}) with $E=1/(2+2K)$.
Although the quantitative differences between the normalized frequencies $F_0(x=+1)$ computed for the event-based corpuscular model
and quantum theory
are larger if the systematic procedure for changing $x$ is used instead of the random procedure, the data obtained with
the random procedure for changing $x$ might be more useful for comparing with the outcomes of
laboratory experiments, as discussed in the next section.

Summarizing: In order to see a difference between the interference patterns predicted by quantum theory and the event-based corpuscular models,
a key factor in the proposed experiment is that the detection events are associated with the value of $x$ at the time of the detection event.
If the detection events are grouped according to the value of $x$, the frequencies of events at detector $D_0$ as obtained
from the event-based corpuscular model are given by Eqs.~(\ref{mzi6a}) and (\ref{mzi6b}). Note
that the difference with Eq.~(\ref{mzi2c}) is only in the prefactors ($E/2$ and $(1-E)/2$ with $0\le E\le1$ instead of 1/2)
which depend on the details of the sequence of $x$-values.

\section{Discussion}

As already mentioned, quantum theory gives an accurate description of the statistics of an experiment
in which the procedure of preparing the particles before they are detected does not change during the experiment.
As the experiment that we propose can be performed such that this condition is not satisfied,
it is of interest to perform this experiment and verify that it agrees with the quantum theoretical prediction.
If the proposed experiment would show deviations from
the quantum theoretical prediction, this finding does not refute quantum theory as such: It provides experimental evidence
that quantum theory cannot be applied to statistical experiments in which the procedure of preparing the particles
before they are detected changes in the course of the experiment.

The event-based corpuscular model~\cite{RAED05d,MICH11a} operates on a level that quantum theory has nothing to say about
and it can easily cope with a preparation procedure that changes with each particle ($K=1$).
As this model reproduces the results of quantum theory under the condition that the
preparation procedure is fixed ($K$ and $N$ large)~\cite{RAED05d,MICH11a}, conventional quantum optics experiments cannot refute
the event-based corpuscular model.
However, as Figs.~\ref{fig3} and \ref{fig4} show,
the proposed MZI experiment with a phase difference alternating between $\phi_0$ and $\phi_0+\pi/2$ (see Fig.~\ref{fig3})
or with a phase difference randomly taking the values $\phi_0$ and $\phi_0+\pi/2$ (see Fig.~\ref{fig4}),
can discriminate between quantum theory and the event-based corpuscular model~\cite{RAED05d,MICH11a}
if the detection events are associated with the value of $x$ at the time of the detection event,
at least in principle.
Recall that if the detection events are not grouped according to the value of $x$ at the time of detection,
both quantum theory and the event-based corpuscular model yield the same interference pattern (see Fig.~\ref{fig2}).

To appreciate the subtilities that are involved, it is necessary to
recognize that there are other experiments in which the preparation procedure is not fixed in time
and for which we do not expect the predictions of quantum theory to deviate from the experimental results,
independent of the pace at which the preparation procedure changes.

As an example, consider Wheeler's delayed choice experiment with single-photons~\cite{JACQ07}.
In this experiment, the random choice between the open and closed configuration of the interferometer
with each passage of a photon does not affect the agreement of the experimental observations with predictions of quantum theory~\cite{JACQ07}.
The reason is that a passage of a photon in the open configuration has no causal effect on
the passage of a photon in the closed configuration.
As the event-based corpuscular approach reproduces the results of quantum theory for Wheeler's delayed choice experiment~\cite{ZHAO08b}
this experiment~\cite{JACQ07} cannot be used to refute the event-based corpuscular model.

The experiment that we propose in this paper is fundamentally different
from e.g. Wheeler's delayed choice experiment with photons in that
the second beam splitter, being the physical cause for interference to occur at all,
is present at all times and that, in a corpuscular picture,
the physical state of a beam splitter may change with each photon passing through it.

\begin{figure}[t]
\begin{center}
\includegraphics[width=11cm]{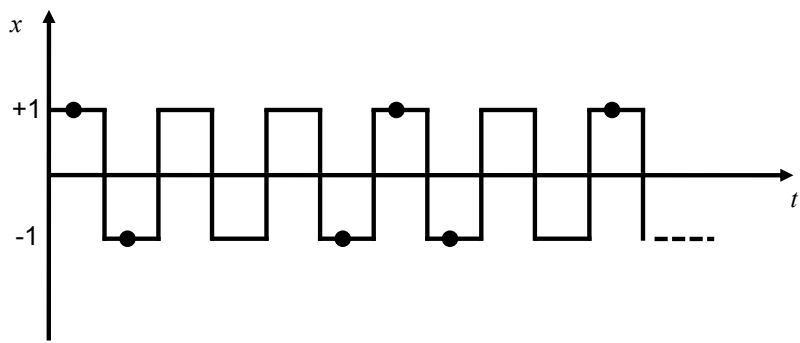}
\caption{%
In the realization of the proposed experiment the variable $x$, taking the values +1 and -1, can be changed alternately in time $t$ at a given fixed rate.
The rate at which the photons (solid circles) are emitted is assumed to be lower than the rate at which $x$ is changed.
Assuming that it is uncertain whether the source emits a photon with each trigger pulse, this experiment is similar to the case where
$x$ is changed according to the random procedure with $K=1$.
}
\label{fig5}
\end{center}
\end{figure}

\section{Realization}

We now address some issues that become relevant when the proposed experiment
is performed in practice.
Essential for the proposed experiment to refute the event-based corpuscular model
or to show the aforementioned limitation of quantum theory is that the rate at which
photons are emitted is lower than the rate at which the time-of-flight in the upper arm of the interferometer
(see Fig.~\ref{fig1}) is being switched between two different values.
Assuming that there is uncertainty about whether or not the source emits a photon
and assuming that the frequency of these pulses
is incommensurate with the frequency with which $x$ changes,
to describe the experiment we may use the model
in which $x$ is changed according to the random procedure with $K=1$, see Fig.~\ref{fig5}.
We emphasize that for the proposed experiment to be successful,
the time-of-flight of a photon from the source to detector should be much less
than the time between changes of $x$ such that there is a one-to-one
correspondence between the value of $x$ and the photon (independent of whether
it is actually detected).
Equally essential is that the procedure to change the time-of-flight of the particles traveling in the upper
arm of the MZI does not alter the particle's direction towards the second beam splitter.

Refuting the event-based corpuscular model~\cite{RAED05d,MICH11a} or to demonstrate the aforementioned limitation of quantum theory
by an experiment will be a real challenge.
The central issue is to collect and analyze the experimental data properly.
To see this, consider the expression for the normalized frequency of events on output channel 0.
In general, that is for $\phi_1(x=+1) \mathrm{mod} 2\pi=0$ and $\phi_1(x=-1)  \mathrm{mod} 2\pi=\delta$,
the event-based corpuscular model predicts
\begin{eqnarray}
{\widetilde I}_{0}(x=+1)&=&\frac{1-E}{2}\sin^2\frac{\phi_0}{2}+\frac{E}{2}\sin^2\frac{\phi_0-\delta}{2}
=\frac{1-\Delta\cos(\phi_0-\psi)}{4}
,
\label{mzi10}
\end{eqnarray}
where $\psi=\arctan(E\sin\delta/(1-E+E\cos\delta))$ and
$\Delta=(2E^2-2E+1+2E(1-E)\cos\delta)^{1/2}$.
From Eq.~(\ref{mzi10}) it follows directly that a least-square fit of a sinusoidal function to the data produced
by the event-based corpuscular model could lead to the conclusion that, independent of the values of $E$ and $\delta$,
this data is described by quantum theory, albeit with a reduced visibility ($|\Delta|<1$).
Thus, this naive procedure to analyse data of single-photon interference experiments
cannot lead to a refutation of the event-based corpuscular model nor can it be used
the test the applicability of quantum theory to event-based experiments.
However, the proposed experiment may be carried out such that there is a chance that
the event-based corpuscular model~\cite{RAED05d,MICH11a} can be refuted and/or these limitations of quantum theory can be demonstrated.

\begin{figure}[t]
\begin{center}
\includegraphics[width=14cm]{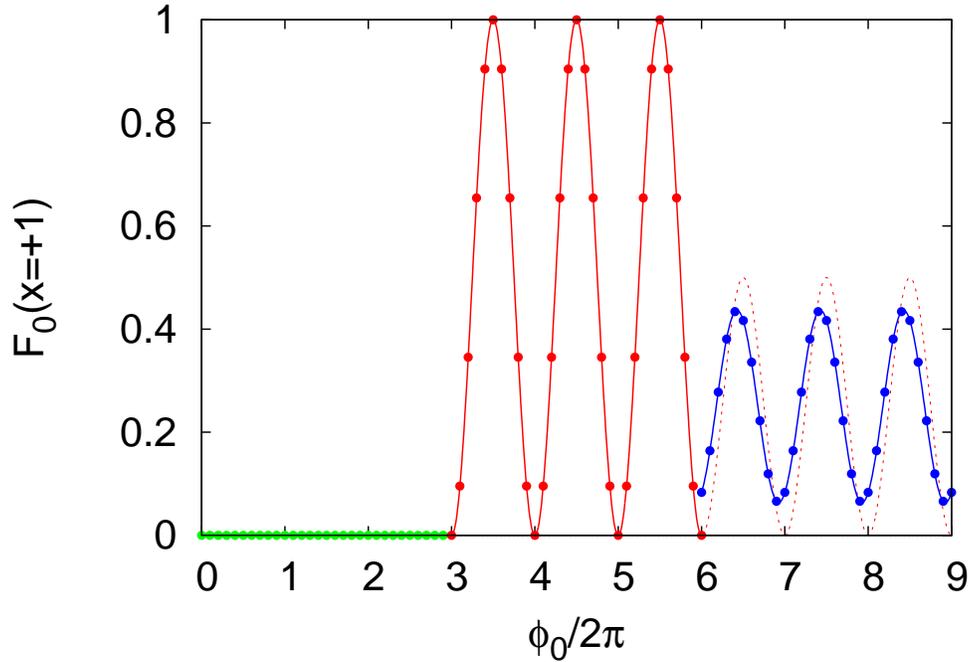}
\caption{%
Results of the normalized frequency $F_0(x=+1)$ in a three-staged MZI experiment
in which the detection events are grouped according to the value of $x$.
First stage ($0\le\phi_0/2\pi< 3$): $x=-1$ fixed.
Second stage ($3\le\phi_0/2\pi< 6$): $x=+1$ fixed.
Third stage ($6\le\phi_0/2\pi< 9$): $x$ is changed according to the systematic procedure with $K=1$.
For each value of $\phi_0$, $N=10^6$ input events were generated and the model parameter $\alpha =0.99$.
Symbols denote the simulation results.
The solid lines are given by Eq.~(\ref{mzi2d}) for the first two stages and by Eq.~(\ref{mzi6a}) with $E=0.33$ for the third stage.
The dotted line is given by quantum theory (Eq.~(\ref{mzi2d}) for stages one and two and Eq.~(\ref{mzi2a}) for stage three.
}
\label{fig6}
\end{center}
\end{figure}

Specifically, for each pulse applied to the single photon source (labeled by the subscript $i$),
the experiment should collect the triples $\{x_i,d_{0,i},d_{1,i}\}$ for $i=1,\dots,N,N+1,\ldots,2N,2N+1,\ldots,3N$
where $d_{k,i}=1$ if detector $D_k$, $k=0,1$ fired (within a properly chosen time window)
and $d_{k,i}=0$ otherwise.
Note that recording both $d_{0,i}$ and $d_{1,i}$ is required for ensuring the single-particle character of the
experiment~\cite{GRAN86}.
For each value of $\phi_0$, in the first stage (the first $N$ pulses),
$x=-1$ is kept fixed while in the second stage of $N$ pulses $x=+1$ kept fixed.
Finally, to mimic a random sequence of $x$-values, in the third stage of $N$ pulses
$x$ should change much faster than the pulse rate at which single photons are emitted.
Assuming that the MZI is stable enough to allow a sufficient amount
of triples to be collected and that the photon flux during the three stages is the same,
comparison of the number of detection counts of the first and second stage
with the one of the third stage,
should or should not (if quantum theory applies) reveal a significant change
in the detection counts (see Fig.~\ref{fig6}).
In other words, performing these three stages in one experimental run should allow one to see a reduction in visibility and a
shift of the sinusoidal curve in the stage in which $x$ changes with respect to the two other stages in which $x$ is fixed.
In experiment this staged procedure may be necessary to compare the reduced visibility (due to experimental limitations)
for the cases with fixed and varying $x$ (which according to quantum theory should all be the same).
Changing the order of the stages and repeating the experiment should provide some
information about the reproducibility of the experimental data.

It will not have escaped the reader that we have not made any assumption about the
efficiency of detecting the photons.
Although for photons this efficiency may be quite low~\cite{HADF09},
this should not affect the conclusions that can be drawn from the
experimental data as long as this data is not contaminated by a significant fraction
of dark counts.
The dark counts may be reduced by using a source emitting pairs of photons in different directions and
by correlating the detection times of the photons detected on detector $D_0$ or $D_1$ placed behind the MZI with those detected on the
detector $D$ placed on the other side of the source (see Fig.~\ref{fig1}).

Although our proposal has been formulated in terms of single-photon experiments,
it should be evident that, at least in theory, one can replace ``photon'' by ``neutron''
without altering the conclusions.
In fact, a neutron experiment which resembles the modified MZI experiment we propose here has been performed~\cite{RAUC84},
but the switching of the conditions was not correlated with the detection events.

We hope that our proposal will stimulate experimenters to take up the challenge to determine
the extent to which quantum theory provides a description of event-based processes that goes beyond statistical averages
or to refute event-based corpuscular models that,
without invoking any concept of quantum theory, reproduce the statistical results of quantum theory.

\section{Conclusions}
We have proposed a single-photon Mach-Zehnder interferometer experiment in which the preparation procedure of the photons in the Mach-Zehnder interferometer
(before detection) is changing in time.
Given
\begin{itemize}
\item[(i)]{the general belief that quantum theory can be used to describe all single-photon experiments,}
\item[(ii)]{the fact that quantum theory gives an accurate description of the statistics of an experiment in which the procedure of preparing the particles
before they are detected does not change during the experiment,}
\item[(iii)]{the fact that the frequency distributions produced by the event-based corpuscular models cannot be distinguished from those predicted by quantum theory
for the single-photon experiments performed so far,}
\item[(iv)]{the fact that the interference patterns of the event-based corpuscular model for the proposed experiment do not agree with those predicted by
quantum theory,}
\end{itemize}
makes this an interesting experiment to be carried out.

The Mach-Zehnder interferometer experiment that we propose has a phase difference alternating between $\phi_0$ and $\phi_0-\delta$ or
randomly taking the values $\phi_0$ and $\phi_0-\delta$ depending on the value of a variable $x\in\{-1,+1\}$.
The variable $x$  may change before the photon enters the MZI but not
during the passage of the photon through the MZI.
The value of $x$ is always known and certain.
If $x$ takes a fixed value during an experimental run then quantum theory and the event-based corpuscular model give the
same interference patterns.
If $x$ is changed (randomly or systematically) and if the detection events are not grouped according to the values that $x$ takes during the
experimental run, then the results predicted by quantum theory and those produced by the event-based corpuscular model also agree.
However, if $x$ is changed and if the detection events are grouped according to the values taken by $x$ then differences appear between the interference patterns
predicted by quantum theory and those produced by the event-based model.
Indeed, quantum theory predicts an interference pattern that is independent of the
sequence of $x$-values whereas
the event-based model shows a reduced visibility and a shift of the interference pattern depending on the
sequence of $x$-values.
Because of the experimental challenges to observe changes in the interference patterns we propose a three-stage experiment:
One stage with $x=-1$ fixed, one stage with $x=+1$
fixed and one stage in which $x$ changes.
We also suggest to interchange the order of the stages in order to study the reproducibility of the experimental outcomes.
To head off possible misunderstandings: If a deviation from a quantum theoretical prediction is observed
this finding would not prove quantum theory wrong but instead would indicate that quantum theory
does not describe the proposed experiment.

Therefore, the key question is: Which interference patterns are produced by a real laboratory experiment?

\section{Acknowledgement}
We thank F.~Jin and K.~De Raedt for discussions.
This work is partially supported by NCF, the Netherlands,
by a Grant-in-Aid for Scientific Research on Priority Areas,
and the Next Generation Super Computer Project, Nanoscience Program from MEXT, Japan.

\bibliographystyle{jpsj}
\bibliography{c:/d/papers/epr11}

\begin{thebibliography}{10}

\bibitem{BORN64}
M.~Born and E.~Wolf: {\em {Principles of Optics}} (Pergamon, Oxford, 1964).

\bibitem{GRAN86}
P.~Grangier, G.~Roger, and A.~Aspect: Europhys. Lett. {\bfseries 1} (1986) 173.

\bibitem{KOCH67}
C.~A. Kocher and E.~D. Commins: Phys. Rev. Lett. {\bfseries 18} (1967) 575 .

\bibitem{GARR09}
J.~C. Garrison and R.~Y. Chiao: {\em {Quantum Optics}} (Oxford University
  Press, Oxford, 2009).

\bibitem{RAED05d}
H.~{De Raedt}, K.~{De Raedt}, and K.~Michielsen: Europhys. Lett. {\bfseries 69}
  (2005) 861 .

\bibitem{RAED05b}
K.~{De Raedt}, H.~{De Raedt}, and K.~Michielsen: Comp. Phys. Comm. {\bfseries
  171} (2005) 19 .

\bibitem{MICH11a}
K.~{Michielsen}, F.~Jin, and H.~{De Raedt}: J. Comp. Theor. Nanosci. {\bfseries
  8} (2011) 1052 .

\bibitem{JACQ07}
V.~Jacques, E.~Wu, F.~Grosshans, F.~Treussart, P.~Grangier, A.~Aspect, and
  J.-F. Roch: Science {\bfseries 315} (2007) 966.

\bibitem{ZHAO08b}
S.~{Zhao}, S.~{Yuan}, H.~{De Raedt}, and K.~Michielsen: Europhys. Lett.
  {\bfseries 82} (2008) 40004.

\bibitem{MICH10a}
K.~Michielsen, S.~Yuan, S.~Zhao, F.~Jin, and H.~{De Raedt}: Physica E
  {\bfseries 42} (2010) 348 .

\bibitem{SCHW99}
P.~D.~D. Schwindt, P.~G. Kwiat, and B.-G. Englert: Phys. Rev. A {\bfseries 60}
  (1999) 4285 .

\bibitem{JIN10c}
F.~{Jin}, S.~{Zhao}, S.~{Yuan}, H.~{De Raedt}, and K.~{Michielsen}: J. Comp.
  Theor. Nanosci. {\bfseries 7} (2010) 1771.

\bibitem{JACQ05}
V.~Jacques, E.~Wu, T.~Toury, F.~Treussart, A.~Aspect, P.~Grangier, and J.-F.
  Roch: Eur. Phys. J. D {\bfseries 35} (2005) 561.

\bibitem{JIN10b}
F.~{Jin}, S.~{Yuan}, H.~{De Raedt}, K.~{Michielsen}, and S.~Miyashita: J. Phys.
  Soc. Jpn. {\bfseries 79} (2010) 074401.

\bibitem{ZHAO08a}
S.~{Zhao} and H.~{De Raedt}: J. Comp. Theor. Nanosci. {\bfseries 5} (2008) 490
  .

\bibitem{AGAF08}
I.~N. Agafonov, M.~V. Chekhova, T.~S. Iskhakov, and A.~N. Penin: Phys. Rev. A
  {\bfseries 77} (2008) 053801.

\bibitem{JIN10a}
F.~{Jin}, H.~{De Raedt}, and K.~{Michielsen}: Commun. Comput. Phys. {\bfseries
  7} (2010) 813 .

\bibitem{RAED05c}
H.~{De Raedt}, K.~{De Raedt}, and K.~Michielsen: J. Phys. Soc. Jpn. Suppl.
  {\bfseries 76} (2005) 16 .

\bibitem{MICH05}
K.~Michielsen, K.~{De Raedt}, and H.~{De Raedt}: J. Comput. Theor. Nanosci.
  {\bfseries 2} (2005) 227 .

\bibitem{ASPE82a}
A.~Aspect, P.~Grangier, and G.~Roger: Phys. Rev. Lett. {\bfseries 49} (1982) 91
  .

\bibitem{ASPE82b}
A.~Aspect, J.~Dalibard, and G.~Roger: Phys. Rev. Lett. {\bfseries 49} (1982)
  1804 .

\bibitem{WEIH98}
G.~Weihs, T.~Jennewein, C.~Simon, H.~Weinfurther, and A.~Zeilinger: Phys. Rev.
  Lett. {\bfseries 81} (1998) 5039 .

\bibitem{RAED06c}
K.~{De Raedt}, K.~Keimpema, H.~{De Raedt}, K.~Michielsen, and S.~Miyashita:
  Euro. Phys. J. B {\bfseries 53} (2006) 139 .

\bibitem{RAED07a}
H.~{De Raedt}, K.~{De Raedt}, K.~Michielsen, K.~Keimpema, and S.~Miyashita: J.
  Phys. Soc. Jpn. {\bfseries 76} (2007) 104005.

\bibitem{RAED07b}
K.~{De Raedt}, H.~{De Raedt}, and K.~Michielsen: Comp. Phys. Comm. {\bfseries
  176} (2007) 642 .

\bibitem{RAED07c}
H.~{De Raedt}, K.~{De Raedt}, K.~Michielsen, K.~Keimpema, and S.~Miyashita: J.
  Comp. Theor. Nanosci. {\bfseries 4} (2007) 957 .

\bibitem{RAED07d}
H.~{De Raedt}, K.~Michielsen, S.~Miyashita, and K.~Keimpema: Euro. Phys. J. B
  {\bfseries 58} (2007) 55 .

\bibitem{ZHAO08}
S.~{Zhao}, H.~{De Raedt}, and K.~Michielsen: Found. of Phys. {\bfseries 38}
  (2008) 322 .

\bibitem{TRIE11}
B.~{Trieu}, K.~{Michielsen}, and H.~{De Raedt}: Comp. Phys. Comm. {\bfseries
  182} (2011) 726.

\bibitem{COMPPHYS}
{\url{http://www.compphys.net/}}.

\bibitem{MZI08}
\url{http://demonstrations.wolfram.com/EventByEventSi}
  \url{mulationOfTheMachZehnderInterferometer/}.

\bibitem{DS08}
\url{http://demonstrations.wolfram.com/EventByEventSi}
  \url{mulationOfDoubleSlitExperimentsWithSinglePhoto/}.

\bibitem{HOME97}
D.~Home: {\em {Conceptual Foundations of Quantum Physics}} (Plenum Press, New
  York, 1997).

\bibitem{BALL03}
L.~E. Ballentine: {\em {Quantum Mechanics: A Modern Development}} (World
  Scientific, Singapore, 2003).

\bibitem{FEYN85}
R.~P. Feynman: {\em QED - The Strange Theory of Light and Matter} (Princeton
  University Press, 1985).

\bibitem{HADF09}
R.~H. Hadfield: Nature Photonics {\bfseries 3} (2009) 696 .

\bibitem{RAUC84}
H.~Rauch and J.~Summhammer: Phys. Lett. A {\bfseries 104A} (1984) 44 .

\end{thebibliography}

\end{document}